\documentclass[aps,twocolumn,noshowkeys,superscriptaddress,noshowpacs]{revtex4}

\usepackage{dcolumn}
\usepackage{amsmath}
\usepackage{bm}
\usepackage{nicefrac}
\usepackage{graphicx}
\usepackage{siunitx}
\usepackage{epsfig}
\usepackage{epstopdf}
\usepackage{xcolor}
\begin{document}

\title{Room-temperature spin glass behavior in zinc ferrite epitaxial thin films}	
\date{\today}

\author{Julia Lumetzberger}
\author{Verena Ney}
\affiliation{Johannes Kepler University Linz, Institute for Semiconductor and Solid State Physics, Altenberger Strasse 69, 4040 Linz, Austria}
\author{Anna Zhakarova}
\author{Nieli Daffe}
\affiliation{Swiss Light Source (SLS), Paul Scherrer Institut, 5232 Villigen PSI, Switzerland}
\author{Daniel Primetzhofer}
\affiliation{Department of Physics and Astronomy, $\dot{A}$ngstr{\"o}m Laboratory,
	Uppsala University, Box 516, SE-751 20 Uppsala, Sweden}
\author{Andreas Ney}
\email{andreas.ney@jku.at}  
\affiliation{Johannes Kepler University Linz, Institute for Semiconductor and Solid State Physics, Altenberger Strasse 69, 4040 Linz, Austria}

\begin{abstract}
Zinc ferrite (ZnFe$_{\text{2}}$O$_{\text{4}}$) epitaxial thin films were grown by reactive magnetron sputtering on MgAl$_{\text{2}}$O$_{\text{4}}$ and Al$_{\text{2}}$O$_{\text{3}}$ substrates varying a range of preparation parameters. The resulting structural and magnetic properties were investigated using a range of experimental techniques confirming epitaxial growth of ZnFe$_{\text{2}}$O$_{\text{4}}$ with the nominal stoichiometric composition and long range magnetic order at and above room temperature. The main preparation parameter influencing the temperature $T_{\text{f}}$ of the bifurcation between $M(T)$ curves under field cooled and zero-field cooled conditions was found to be the growth rate of the films, while growth temperature or the Ar:O$_2$ ratio did not systematically influence $T_{\text{f}}$. Furthermore $T_{\text{f}}$ was found to be systematically higher for MgAl$_{\text{2}}$O$_{\text{4}}$ as substrate and $T_{\text{f}}$ extends to above room temperature. While in some samples $T_{\text{f}}$ seems to be more likely correlated with superparamagentism, the highest $T_{\text{f}}$ occurs in ZnFe$_{\text{2}}$O$_{\text{4}}$ epitaxial films where experimental signatures of magnetic glassiness can be found. Element-selective X-ray magnetic circular dichroism measurements aim at associating the magnetic glassiness with the occurrence of a different valence state and lattice site incorporation of Fe pointing to a complex interplay of various competing magnetic interactions in ZnFe$_{\text{2}}$O$_{\text{4}}$.
\end{abstract}

\maketitle

\section{Introduction}

Zinc ferrite (ZnFe$_{\text{2}}$O$_{\text{4}}$) belongs to the crystallographic group of normal spinels of the form AB$_\text{2}$O$_\text{4}$ where in the ideal case the A-cation (Zn$^{2+}$) exclusively occupies the tetrahedral (Td) lattice sites as Zn$^{2+}_{Td}$\ while the B-cation (Fe$^{3+}$) is found on the octahedral (Oh) sites as Fe$^{3+}_{Oh}$. The magnetic properties of zinc ferrite have been under investigation for quite some time revealing a rather complex situation. Early studies of the bulk material report antiferromagnetic (AFM) order with a very low N\'eel temperature of \SI{9}{\kelvin} \cite{HaC53,HaC56}. Later-on it was demonstrated by magnetic neutron scattering experiments on ZnFe$_{\text{2}}$O$_{\text{4}}$ single crystals that even in perfect crystals geometrical frustration leads to an unusual magnetic behavior \cite{KTK03}. In particular, it was pointed out, that the Fe$^{3+}_{Oh}$ sublattice can be regarded to be similar to various pyrochlores or Laves phases which are known for their intrinsic geometrical frustration \cite{KTK03}. The situation becomes even more complex when defects such as inversion are considered which are expected to occur, in particular, in thin films of ZnFe$_{\text{2}}$O$_{\text{4}}$. A partial inversion in ZnFe$_{\text{2}}$O$_{\text{4}}$ has the stoichiometric formula of [Zn$_{1-\delta}$Fe$_{\delta}$]$_{Td}$[Zn$_{\delta}$Fe$_{2-\delta}$]$_{Oh}$O$_4$, where $\delta$ denotes the degree of inversion. In the ideal case of $\delta=0$, i.\,e., no inversion, there is only the weak AFM superexchange interaction between Fe$^{3+}_{Oh}$ which is usually denoted as J$_{\text{BB}}$ \cite{VAN09} plus the geometrical frustration mentioned before \cite{KTK03}. J$_{\text{BB}}$ can be held responsible for the AFM order at low temperatures in bulk single crystals. For a finite degree of inversion there is an additional, much stronger AFM superexchange interaction J$_{\text{AB}}$ between Fe$^{3+}$ on Td (also called A-site) and Oh (or B-site) sites, i.\,e., Fe$^{3+}_{Td}$ and Fe$^{3+}_{Oh}$ \cite{VAN09}, which leads to ferrimagnetism for incomplete inversion \cite{ZGS20}. The additional J$_{\text{AA}}$ exchange between the Fe$^{3+}_{Td}$ is the weakest \cite{RTG11}. However, if the additional Fe$^{3+}_{Td}$ is not compensated by Zn$^{2+}_{Oh}$, i.\,e., if there is some degree of deviation from the ideal stoichiometry of ZnFe$_{\text{2}}$O$_{\text{4}}$, some finite amount of Fe$^{2+}_{Oh}$ has to form because of charge neutrality. This results in an additional double exchange (DE) over the Oh (or B-) sites J$_{\text{BB}}^{\text{DE}}$ between Fe$^{3+}_{Oh}$ and Fe$^{2+}_{Oh}$, which results in spin canting in magnetite \cite{VAN09} or in non-stoichiometric ZnFe$_{\text{2}}$O$_{\text{4}}$ \cite{ZGS20}. In many cases there are reports on some finite degree of inversion in ZnFe$_{\text{2}}$O$_{\text{4}}$ and an upper limit of $\delta = 0.6$ has been found by the analysis of the magnetic moment of the Fe \cite{YTK01}, X-ray absorption spectroscopy (XAS) \cite{NFT07} or via Rietveld refinement of X-ray diffraction (XRD) data \cite{CPL19}. Therefore, the magnetic order in ZnFe$_{\text{2}}$O$_{\text{4}}$ can be expected to be highly complex, especially for thin films, where the presence of various kinds of defects such inversion and/or off-stoichiometry can be expected. 

The growth of zinc ferrite in thin film form is motivated by a range of possible applications such as gas sensors, photo catalytic disinfection or other photo-catalytic applications, see \cite{CPL19,PTK17} and Refs. therein. Besides that, zinc ferrite thin films can also be considered as an interesting semiconducting material in spintronics with tunable magnetic properties \cite{MYN07,CSZ08,VAN09,RTG11,TAS13}. A variety of reports of different types of magnetic order in zinc ferrite can be found throughout the literature ranging from ferro(i)magnetic \cite{VAN09,SuS09,RTG11,TAS13,GJF14,BBL18,ZSE20}, to superparamagnetic \cite{CSH95,MYN07,NFT07,KLL17,MOR21} and to spin glass behavior \cite{CPL19,YTK01,NFT05,HMH11}. Note, that in some cases superparamagnetism with interparticle interactions is associated with a so-called cluster-glass behavior \cite{NFT05,NFT07,YTK01}. However, only a few studies report on characteristic experimental signatures of magnetic glassiness \cite{CPL19,NFT05,YTK01,HMH11}, while others report only temperature-dependent magnetization [$M(T)$] measurements under different field-cooling conditions which however could also be associated with superparamagnetism, e.\,g. \cite{NFT07}. Finally, the control of defects and thus the magnetic properties was reported to be experimentally achievable by varying different preparation parameters, i.\,e., oxygen partial pressure \cite{CSZ08,RTG11,CSH95,ZSE20,MOR21}, stoichiometric composition \cite{VAN09}, post-growth thermal treatment \cite{NFT05,KLL17} or deposition rate \cite{YTK01}.

Similar to the reported types of magnetism in zinc ferrite, also the techniques for sample preparation span a wide range from pulsed laser deposition (PLD) \cite{MYN07,CSZ08,VAN09,RTG11,TAS13,KLL17,BBL18,YTK01,ZSE20}, over reactive magnetron sputtering (RMS) \cite{CSH95,NFT05,NFT07,SuS09,GJF14,MOR21} for thin film growth, to ball milling \cite{CPL19} and solid state reaction \cite{HMH11} for bulklike samples. Likewise, a range of different substrates has been used for thin film growth amongst them are MgAl$_{\text{2}}$O$_{\text{4}}$ \cite{CSZ08}, c-plane Al$_{\text{2}}$O$_{\text{3}}$ \cite{MYN07,YTK01}, a-plane Al$_{\text{2}}$O$_{\text{3}}$ \cite{BBL18}, SrTiO$_\text{3}$ \cite{TAS13,KLL17,ZSE20}, MgO \cite{MYN07,VAN09,RTG11}, Si(001) and Si(111) \cite{GJF14,MOR21} and glass substrates \cite{CSH95,NFT05,NFT07,SuS09}. It is remarkable, that spin glass behavior has mainly been reported for bulk ZnFe$_{\text{2}}$O$_{\text{4}}$ \cite{KTK03} or bulklike nanopowders \cite{CPL19,HMH11} while for thin film samples mostly a cluster glass is inferred \cite{NFT05,NFT07,YTK01}. Among the reports of cluster glass behavior only one is based on thin film growth by PLD on single crystalline substrates \cite{YTK01}, while the others rely on sputtered polycrystalline ZnFe$_{\text{2}}$O$_{\text{4}}$ samples \cite{NFT05,NFT07} making a larger amount of defects expectable, e.\,g., due to an intrinsically large number of grain boundaries. Finally, also the temperature range, where magnetic glassiness is observed ranges from below 20~K for the single crystalline ZnFe$_{\text{2}}$O$_{\text{4}}$ in \cite{KTK03,HMH11}, over around 100~K for the annealed ZnFe$_{\text{2}}$O$_{\text{4}}$ nanopowder in \cite{CPL19} up to 300~K for the PLD grown ZnFe$_{\text{2}}$O$_{\text{4}}$ epitaxial films at deposition rates above 3~nm/s which drops down to below 100~K for rates below 2~nm/s \cite{YTK01}. 

Here, we report on epitaxial thin film growth of ZnFe$_{\text{2}}$O$_{\text{4}}$ by RMS on two different substrates namely MgAl$_{\text{2}}$O$_{\text{4}}$ and c-plane Al$_{\text{2}}$O$_{\text{3}}$. Various preparation parameters have been varied in order to control the formation of defects in a systematic way for epitaxial thin film samples. In agreement with \cite{YTK01} the most relevant preparation parameter is found to be the deposition rate. For high deposition rates ZnFe$_{\text{2}}$O$_{\text{4}}$ films exhibit spin-glass like behavior up to rather high temperatures on MgAl$_{\text{2}}$O$_{\text{4}}$ substrates which is shifted to lower temperatures on Al$_{\text{2}}$O$_{\text{3}}$ substrates. In contrast, the stoichiometry on ZnFe$_{\text{2}}$O$_{\text{4}}$ is maintained throughout the sample series and also the oxygen partial pressure were found to play a minor role in the resulting magnetic properties. The spin-glass behavior is associated with a significant amount of inversion up to $\delta \sim 0.3$, corroborating earlier reports \cite{CPL19,NFT07}. In addition, a significant magnetic polarization of the Zn cation is found at room temperature by means of element selective magnetometry indicating that the microscopic origin of the magnetic properties of ZnFe$_{\text{2}}$O$_{\text{4}}$ is even more complex. 

\section{Experimental Details}

Zinc ferrite was fabricated using reactive magnetron sputtering (RMS) from an oxide target having the nominal composition of ZnFe$_{\text{2}}$O$_{\text{4}}$. The epitaxial thin films were grown on doubleside polished single crystalline spinel [MgAl$_{\text{2}}$O$_{\text{4}}$(001)] and c-plane sapphire [Al$_{\text{2}}$O$_{\text{3}}$(0001)] substrates in an ultrahigh vacuum (UHV) chamber with a base pressure of $4 \times 10^{-8}$\,\si{\milli\bar} and a working pressure of $4 \times 10^{-3}$\,\si{\milli\bar}. To determine the ideal growth parameters, the deposition temperature was varied from room temperature (RT) to \SI{550}{\degreeCelsius}, the Ar:O$_2$ ratio from $10$\,:\,$0$ to $10$\,:\,$0.5$, and the sputtering power from \SIrange{20}{100}{\watt}, which corresponds to a growth rate from 0.36 to 3.69~nm/min. The nominal thickness is kept at 40~nm and is controlled via a quartz crystal microbalance which is at room temperature so that the actual thickness of most of the films is by 10-20\% lower because of the elevated temperature of the substrate during growth. The structural properties of the films were investigated by X-ray diffraction (XRD) measurements with a \textit{Pananalytical X'Pert MRD} recording $\omega - 2 \theta$ scans and symmetric as well as asymmetric reciprocal space maps (RSM). The chemical composition was determined by ion beam analysis, i.e. Rutherford backscattering spectrometry (RBS) using a \SI{2}{\mega\electronvolt} He$^{+}$ primary beam at the Tandem Laboratory at Uppsala University. To disentangle the element specific contributions, the spectra were analyzed using the SIMNRA software \cite{May97}. Details of the experimental setup are described elsewhere \cite{MHZ19}. Furthermore, Electron Recoil Detection (ERDA) with a primary ion beam of \SI{36}{\mega\electronvolt} iodine ions was employed to rule out contaminations with light elements like H or C.

The magnetic properties were measured by integral superconducting quantum interference device (SQUID) magnetometry using a \textit{Quantum Design MPMS-XL5} system applying the magnetic field in the film plane. The $M(H)$ curves were recorded in range of $\pm 5$\,\si{\tesla} at \SI{300}{\kelvin} and \SI{2}{\kelvin} and $M(T)$ curves have been recorded from \SI{2}{\kelvin} up to \SI{395}{\kelvin} at 10~mT while warming after a cool down in 5~T (FH), under nominally zero-field cooled conditions (ZFC) as well while cooling down in 10~mT (field cooled, FC). Additionally, waiting time experiments were performed analogous to the ones in \cite{BeK09} by cooling down the sample in zero field and introducing a waiting time $t_{wait}$ at various waiting temperatures $T_{wait}$ which is typically \SI{10000}{\second}. Then a $M(T)$ curve identical to a ZFC curve without waiting time was subsequently recorded. Subtracting these two curves represents a typical waiting time experiment for spin glasses where a so-called ZFC memory---or hole-burning---effect can be seen by a dip in the difference of the magnetization with and without waiting time around $T_{wait}$ \cite{BeK09}. A second memory experiment already used before for ZnFe$_{\text{2}}$O$_{\text{4}}$ in \cite{CPL19} was performed in addition, in which a $M(T)$ curve is recorded under FC conditions with 10~mT and a waiting time $t_{wait}$ at nominally zero field is inserted at several $T_{wait}$ before the FC curve is resumed at 10~mT. Then a subsequent $M(T)$ is recorded in 10~mT while warming. The typical signature of a spin glass in these so-called FC memory experiments is a relaxation of the magnetization at $T_{wait}$ in the FC curve and the presence of an inflection point around $T_{wait}$ in the subsequent $M(T)$ curve while warming \cite{CPL19}. Note, however, that also superparamagnets exhibit similar signatures in the FC memory experiments \cite{SJT05}. All $M(H)$ and $M(T)$ data were corrected for the diamagnetic background of the substrate which was determined from the $M(H)$ curves a high magnetic field at 300~K \cite{SSN11}. The $M(H)$ curves at 2~K for samples grown on MgAl$_{\text{2}}$O$_{\text{4}}$ had to be corrected for an additional paramagnetic contribution which was determined form a $M(H)$ measurement of bare MgAl$_{\text{2}}$O$_{\text{4}}$ from the same batch of samples. In general, zero field conditions are referred to nominally 0.0~mT after the superconducting magnet had been reset (magnet reset option of the MPMS) and any applied magnetic field is afterwards limited to 10~mT; this assures a residual pinned magnetic field of typically 0.1~mT or less \cite{BHH18}. A cooling and heating rate of 1~K/min is used for all $M(T)$ measurements in both FC and ZFC memory experiments. Note, that the typical frequency-dependent ac-susceptibility measurements like in \cite{NFT05,HMH11,YTK01} were not available for the used SQUID.

The element specific magnetic properties have been investigated by x-ray absorption near edge spectroscopy (XANES) and x-ray magnetic circular dichroism (XMCD) measurements which were performed at the Xtreme beamline at the Swiss Light Source (SLS) \cite{PFR12}. The XMCD spectra were recorded at the Fe $L_{3/2}$- and Zn $L_{3/2}$-edges at \SI{300}{\kelvin} under \SI{20}{\degree} grazing incidence in total electron yield. For the Fe-edges the magnetic field was set to 5~T and only the circular polarization has been switched to obtain the XMCD. For the Zn-edges the direction of the magnetic field has been reversed as well to minimize artifacts. The XMCD spectra at the Fe $L_3$ edge are compared to simulations carried out by multiplet ligand field theory using the CTM4XAS package \cite{SdG10}. These simulations have been used before to determine the site occupancy and formal oxidation state of Fe and Ni in nickel ferrites \cite{KMK14} and Zn/Al doped nickel ferrites \cite{LBP20}. For the present work the simulation parameters for Fe$^{2+}_{Oh}$, Fe$^{3+}_{Oh}$, and Fe$^{3+}_{Td}$ are identical to those in \cite{LBP20} and details on the simulations can be found there.

\section{Experimental results}

\begin{figure}[t]
	\centering
		\includegraphics[width=0.47\textwidth]{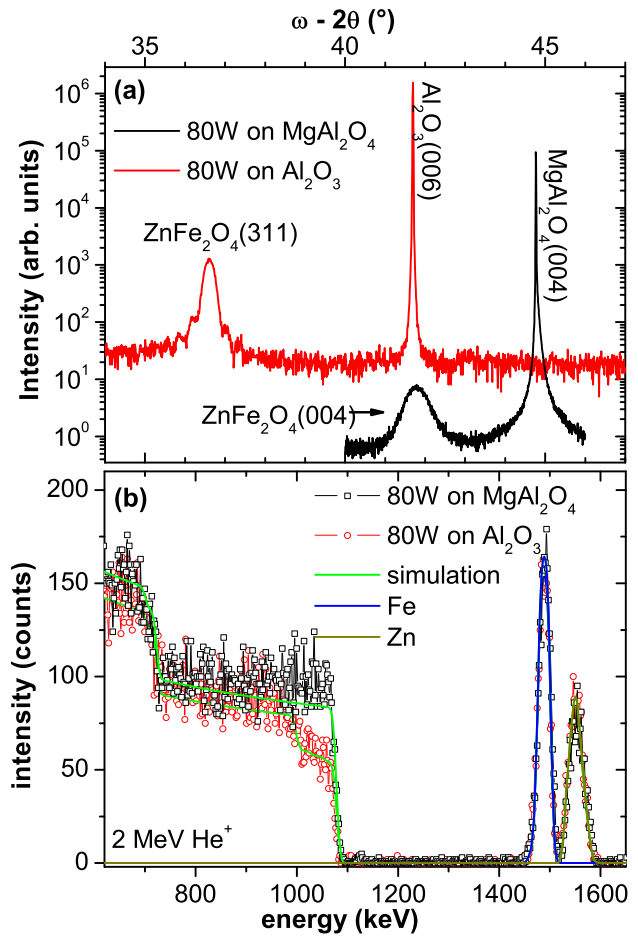}
		\vspace{-0.2cm}
	\caption{(a) Structural characterization by X-ray diffraction: comparison of the symmetric $\omega-2 \theta$ scans of ZnFe$_{\text{2}}$O$_{\text{4}}$ films grown at 80~W on MgAl$_{\text{2}}$O$_{\text{4}}$ (black) and Al$_{\text{2}}$O$_{\text{3}}$ (red). (b) Chemical composition determined by means of RBS spectra recorded with a 2~MeV He$^+$ primary ion beam of ZnFe$_{\text{2}}$O$_{\text{4}}$ films grown at 80~W on MgAl$_{\text{2}}$O$_{\text{4}}$ (black) and Al$_{\text{2}}$O$_{\text{3}}$ (red).}
	\label{fig1}
\end{figure}

The structural properties of the ZnFe$_{\text{2}}$O$_{\text{4}}$ thin films were analyzed by symmetric $\omega-2 \theta$ scans using XRD. Figure \ref{fig1}(a) shows a comparison of the diffractograms of zinc ferrite grown on MgAl$_{\text{2}}$O$_{\text{4}}$ and Al$_{\text{2}}$O$_{\text{3}}$ where the latter is shifted upward for clarity. The samples were grown with a nominal thickness of \SI{40}{\nano\meter} at a substrate temperature of $T_{S}=\,$\SI{450}{\degreeCelsius} with an Ar:O$_2$ ratio of $ 10 $\,:\,$0.5$ and a sputtering power of \SI{80}{\watt}. The XRD scan of the samples grown on MgAl$_{\text{2}}$O$_{\text{4}}$ exhibits a ($004$) reflex at $41.79 \pm 0.09$\,\si{\degree} with a full width at half maximum (FWHM) of $0.6$\,\si{\degree}, corresponding to a perpendicular lattice parameter $a_{\perp} = (8.64 \pm 0.02)$\,\si{\angstrom}. The sample grown on Al$_{\text{2}}$O$_{\text{3}}$ exhibits the ($311$) reflex at \SI{35.61}{\degree} corresponding to a perpendicular lattice parameter of $a_{\perp} = (7.28 \pm 0.02)$\,\si{\angstrom}. For this sample weak Laue oscillations can be seen indicating a smoother growth compared to the sample grown on MgAl$_{\text{2}}$O$_{\text{4}}$. In addition, an asymmetric RSM along the ($\bar{1}\bar{1}5$) plane  has been recorded for a \SI{100}{\nano\meter} sample grown on MgAl$_{\text{2}}$O$_{\text{4}}$ with a sputtering power of \SI{60}{\watt} (not shown). It reveals an in-plane lattice parameter of $a_{\parallel} = (8.32 \pm 0.05)$\,\si{\angstrom} which provides evidence that the film is relaxed, since the film peak does not align with the substrate peak. The reflection from the MgAl$_{\text{2}}$O$_{\text{4}}$ substrate corresponds to a lattice parameter of $a_{\text{sub}} = 8.08$\,\si{\angstrom}, which implies a lattice mismatch of $\sim$\,\SI{4.3}{\percent} with respect to bulk ZnFe$_{\text{2}}$O$_{\text{4}}$ ($a_0 = 8.441$\,\si{\angstrom}) (JCPDS card No. 82-1049). A comparison between various films grown on MgAl$_{\text{2}}$O$_{\text{4}}$ and Al$_{\text{2}}$O$_{\text{3}}$ indicates no significant difference in crystalline quality with similar FWHM despite the change in texture from ($004$) on MgAl$_{\text{2}}$O$_{\text{4}}$ to ($311$) on Al$_{\text{2}}$O$_{\text{3}}$. No other reflexes can be found underlining highly textured growth of ZnFe$_{\text{2}}$O$_{\text{4}}$ for both substrates and the samples are devoid of other crystalline phases. Furthermore, the chemical composition of ZnFe$_{\text{2}}$O$_{\text{4}}$ on both substrates is determined using RBS. In Fig.\,\ref{fig1}(b) the RBS data of the \SI{80}{\watt} sample grown on MgAl$_{\text{2}}$O$_{\text{4}}$ (black squares) is shown in comparison to the sample grown on Al$_{\text{2}}$O$_{\text{3}}$ (red circles). Both ZnFe$_{\text{2}}$O$_{\text{4}}$ films have no deviation from the nominal stoichiometry within the uncertainties of the measurement technique. Finally, the samples are investigated using ERDA to check for contaminations of light elements like H or C but neither element could be detected (not shown). We can therefore conclude that our ZnFe$_{\text{2}}$O$_{\text{4}}$ samples have the nominal stoichiometric composition, grow epitaxially on either substrate and are devoid of a significant amount of secondary phases or contaminants within the detection limits of XRD and ERDA. 

\begin{figure}[t]
	\centering
	\includegraphics[width=0.45\textwidth]{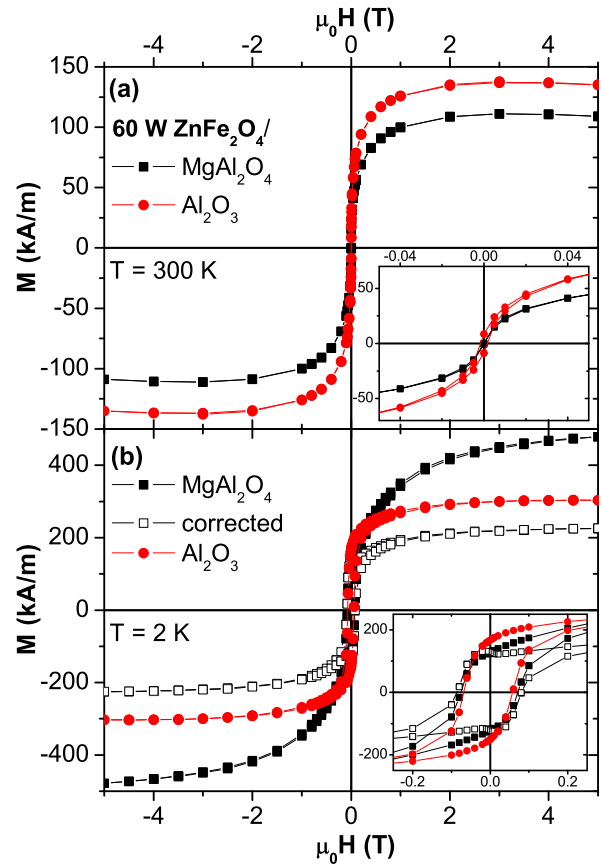}
	\vspace{-0.2cm}
	\caption{SQUID measurements of $M(H)$ curves shown for the 60~W ZnFe$_{\text{2}}$O$_{\text{4}}$ grown on MgAl$_{\text{2}}$O$_{\text{4}}$ (black squares) and Al$_{\text{2}}$O$_{\text{3}}$ (red circles) at (a) 300~K and (b) at 2~K. (a) shows an M(H) curve at RT (b). At 2~K the paramagnetic contribution of the MgAl$_{\text{2}}$O$_{\text{4}}$ substrate has been subtracted (open squares). The insets enlarge the measurements at low fields.}
	\label{fig2}
\end{figure} 

In a first step, the magnetic properties are investigated using standard $M(H)$ curves which are shown in Fig.\ \ref{fig2} recorded at (a) \SI{300}{\kelvin} and (b) \SI{2}{\kelvin} for ZnFe$_{\text{2}}$O$_{\text{4}}$ grown at \SI{60}{\watt} on MgAl$_{\text{2}}$O$_{\text{4}}$ (black squares) and Al$_{\text{2}}$O$_{\text{3}}$ (red circles). ZnFe$_{\text{2}}$O$_{\text{4}}$ grown on Al$_{\text{2}}$O$_{\text{3}}$ has a higher magnetization of $M_{\text{S}} = 130 \pm 15 $\,\si{\kilo\ampere/\meter} compared to growth on MgAl$_{\text{2}}$O$_{\text{4}}$ where $M_{\text{S}} = 110 \pm 15 $\,\si{\kilo\ampere/\meter} at \SI{300}{\kelvin}. For both samples $M_{\text{S}}$ increases to above $200$\,\si{\kilo\ampere/\meter} at \SI{2}{\kelvin}. For the $M(H)$ curves recorded at \SI{2}{\kelvin} for ZnFe$_{\text{2}}$O$_{\text{4}}$ grown on MgAl$_{\text{2}}$O$_{\text{4}}$ the full squares denote the data when only the diamagnetic contribution has been subtracted. Note, that in a previous publication on ZnFe$_{\text{2}}$O$_{\text{4}}$ grown on MgAl$_{\text{2}}$O$_{\text{4}}$ this apparently paramagnetic behavior has been attributed to cationic disorder of the Fe$^{3+}$ in \cite{CSZ08}. However, if a bare MgAl$_{\text{2}}$O$_{\text{4}}$ substrate is measured, one also measures a net-paramagnetic behavior after subtraction of the diamagnetism so that one has to attribute this paramagnetic contribution to the MgAl$_{\text{2}}$O$_{\text{4}}$ substrate itself. The open squares are the data where also the measured paramagnetic background of the bare MgAl$_{\text{2}}$O$_{\text{4}}$ has been subtracted and no obvious paramagnetic contribution of the ZnFe$_{\text{2}}$O$_{\text{4}}$ is visible any more. The insets in Fig.\ \ref{fig2} enlarge the low-field behavior of the $M(H)$ curves. While they are virtually anhysteretic at \SI{300}{\kelvin} a clear hysteresis with a coercive field of $H_{\text{c}}= 80 \pm 10$\,\si{\milli\tesla} is found for ZnFe$_{\text{2}}$O$_{\text{4}}$ films on either substrate. This behavior at 2~K is consistent with most of the ZnFe$_{\text{2}}$O$_{\text{4}}$ films grown on a range of different substrates reported throughout the literature reporting ferro(i)magnetism or superparamagnetism both in terms of magnetization as well as coercive field at low temperatures and clearly rules out the pure antiferromagnetic behavior of bulk ZnFe$_{\text{2}}$O$_{\text{4}}$.

\begin{figure}[t]
	\centering
	\includegraphics[width=0.45\textwidth]{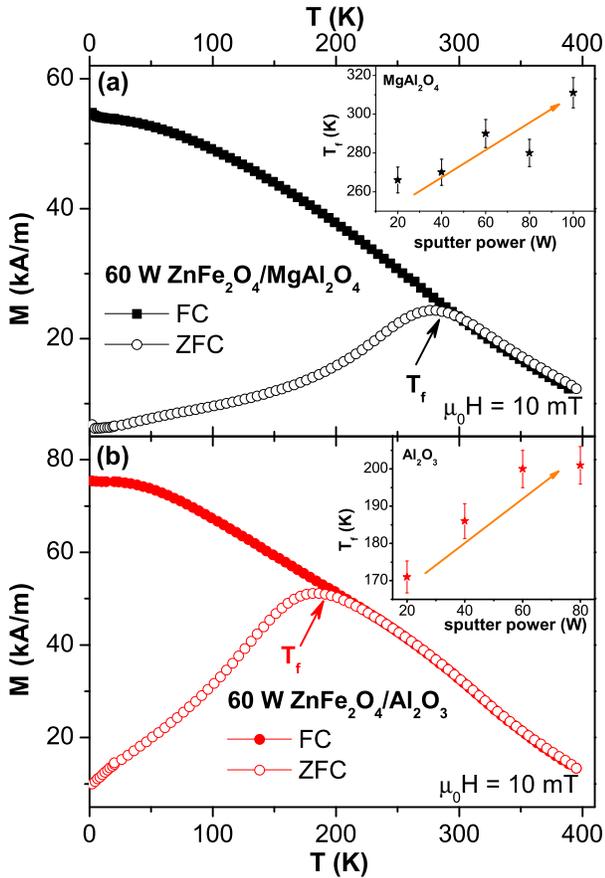}
	\vspace{-0.2cm}
	\caption{$M(T)$ curves recorded at 10~mT under field cooled (FC, full symbols) and after zero-field cooled (ZFC, open symbols) conditions shown for the 60~W ZnFe$_{\text{2}}$O$_{\text{4}}$ grown on (a) MgAl$_{\text{2}}$O$_{\text{4}}$ (black squares) and (b) Al$_{\text{2}}$O$_{\text{3}}$ (red circles) substrates. The insets show the dependence of $T_f$ on the sputtering power for both substrates, respectively.}
	\label{fig3}
\end{figure} 

Figure \ref{fig3} shows the $M(T)$ behavior recorded at \SI{10}{mT} under FC conditions (full symbols) as well as after ZFC conditions (open symbols) for the ZnFe$_{\text{2}}$O$_{\text{4}}$ grown at \SI{60}{\watt} on MgAl$_{\text{2}}$O$_{\text{4}}$ (a) and Al$_{\text{2}}$O$_{\text{3}}$ (b), i.\,e., the identical pair of samples as in Fig.\ \ref{fig2}.  Both samples exhibit a clear bifurcation between the FC and ZFC curves indicating a blocking or spin-freezing peak at a temperature of $T_{\text{f}}= 290$\,\si{\kelvin} for ZnFe$_{\text{2}}$O$_{\text{4}}$/MgAl$_{\text{2}}$O$_{\text{4}}$ (a) and at $T_{\text{f}}= 190$\,\si{\kelvin} for ZnFe$_{\text{2}}$O$_{\text{4}}$/Al$_{\text{2}}$O$_{\text{3}}$ (b). Both \SI{60}{\watt} samples together with the two \SI{80}{\watt} samples shown in Fig.\ \ref{fig1} are part of a sample series where only the sputtering power and thus the deposition rate has been changed while all other growth parameters have been kept constant. In terms of magnetization as well as coercivity all samples from these series show comparable magnetic behavior. The only systematic dependency on the sputtering power is an increase of the measured $T_{\text{f}}$ with increasing sputtering power. The insets in Fig.\ \ref{fig3} show the measured $T_{\text{f}}$ as a function of the sputtering power for ZnFe$_{\text{2}}$O$_{\text{4}}$/MgAl$_{\text{2}}$O$_{\text{4}}$ (a) as well as for ZnFe$_{\text{2}}$O$_{\text{4}}$/Al$_{\text{2}}$O$_{\text{3}}$ (b). Irrespective of the comparable increase with sputtering power the overall values of $T_{\text{f}}$ are systematically lower for the Al$_{\text{2}}$O$_{\text{3}}$ substrate by about \SI{100}{\kelvin}. Note that the obtained $T_{\text{f}}$ for the samples grown on either substrate, are well above the values usually reported for zinc ferrite \cite{KLL17,CPL19,HMH11,ZSE20}. Only in few cases the $T_{\text{f}}$ is found at such elevated temperatures \cite{NFT05,YTK01} and a controllable shift of $T_{\text{f}}$ is only reported in \cite{YTK01} so far; however, the drop in $T_{\text{f}}$ with decreasing deposition rate in \cite{YTK01} is by a factor of two more pronounced compared with the present case. Note, that the power series was grown by varying the sputter power nonmonotonously so that a dependence of $T_{\text{f}}$ on the growth sequence---and thus target degradation---can be ruled out.   

\begin{figure}[t]
	\centering
	\includegraphics[width=0.45\textwidth]{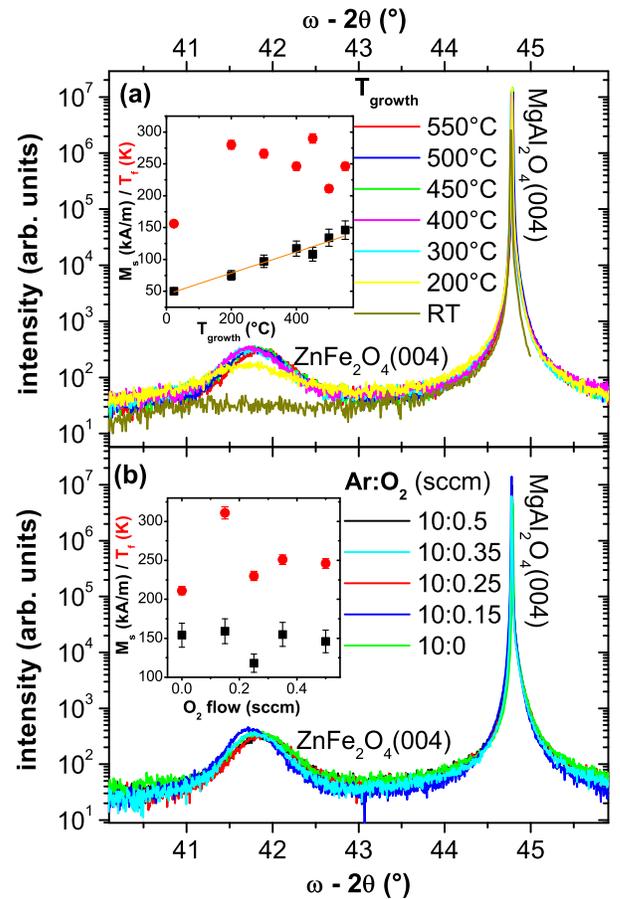}
	\vspace{-0.2cm}
	\caption{(a) Structural and resulting magnetic properties as a function of the growth temperature $T_{growth}$ for ZnFe$_{\text{2}}$O$_{\text{4}}$ grown on MgAl$_{\text{2}}$O$_{\text{4}}$. The dependence of the identical parameters as a function of the Ar:O$_2$ ratio is shown in (b). $M_s$ (black squares) and $T_f$ (red circles) shown in the insets were extracted from SQUID measurements.}
	\label{fig4}
\end{figure} 

In a second step, the dependence of $T_{\text{f}}$ on other growth parameters shall be briefly summarized. Figure \ref{fig4}(a) compiles the sample series as a function of the growth temperature $T_{\text{growth}}$ from RT up to \SI{550}{\degreeCelsius} for ZnFe$_{\text{2}}$O$_{\text{4}}$ grown on MgAl$_{\text{2}}$O$_{\text{4}}$ at a sputtering power of \SI{60}{\watt} and an Ar:O$_2$ ratio of $ 10 $\,:\,$0.5$. The XRD shows no significant changes for $T_{\text{growth}} \geq $ \SI{300}{\degreeCelsius} while the sample at RT appears to be virtually amorphous. The inset shows $M_{\text{S}}$ at 300~K and $T_{\text{f}}$ as determined from SQUID measurements analogous to Figs. \ref{fig2} and \ref{fig3}. $T_{\text{f}}$ is found to be about constant around \SI{250}{\kelvin} with a slight tendency to decrease for higher $T_{\text{growth}}$; the only exception is the amorphous sample at RT where $T_{\text{f}}$ is clearly reduced. In contrast, $M_{\text{S}}$ steadily increases with increasing $T_{\text{growth}}$ which can be taken as an indication for an increasing amount of inversion, i.\,e., of Fe$^{3+}_{Td}$ in analogy with \cite{CPL19,YTK01}. However; for the present sample series this increasing $M_{\text{S}}$ with $T_{\text{growth}}$ has no obvious influence on the observed $T_{\text{f}}$. This trend is opposite to the annealing series in \cite{CPL19}, where the amount of inversion and thus resulting $M_{\text{S}}$ is decreasing with increasing annealing temperatures for nanopowdered ZnFe$_{\text{2}}$O$_{\text{4}}$. Note, that increasing $T_{\text{growth}}$ in epitaxial growth typically leads to a decrease in actual thickness compared to the nominal one which would lead to a decrease in $M_{\text{S}}$ which was calculated from the nominal thickness. On the other hand, this increase in $M_{\text{S}}$ can also be associated with an increase in the order temperature which is in all cases above 400~K and thus beyond the accessible temperature range of the SQUID magnetometer and thus unknown.

A second ZnFe$_{\text{2}}$O$_{\text{4}}$ sample series was grown on MgAl$_{\text{2}}$O$_{\text{4}}$ at a sputtering power of \SI{60}{\watt} and a fixed $T_{\text{growth}}$ of \SI{450}{\degreeCelsius} while varying the Ar:O$_2$ ratio which is compiled in Fig.\ \ref{fig4}(b). The XRD of all samples does not show any significant changes with increasing oxygen content. In the inset the resulting $M_{\text{S}}$ at 300~K and $T_{\text{f}}$ are shown. While $M_{\text{S}}$ is independent on the Ar:O$_2$ ratio within error bars, $T_{\text{f}}$ does not show a conclusive trend, mostly because of a rather high $T_{\text{f}}$ for the sample at Ar:O$_2$ ratio of $ 10 $\,:\,$0.15$. Disregarding this, a faint increase within errorbars may be inferred but there is no pronounced dependence of $T_{\text{f}}$ on the Ar:O$_2$ ratio, especially if this is compared with the dependence on the growth rate shown in Fig.\ \ref{fig3}(a). This finding is rather interesting, since in \cite{YTK01} the growth rate has been associated with a deficiency in oxygen leading to an increase in $T_{\text{f}}$. However, all samples were found to be highly resistive above the G$\Omega$-range (not shown). Therefore, a significant amount of oxygen vacancies can be ruled out for the entire series because the two samples grown without and with maximum oxygen partial pressure have virtually identical physical properties where the high oxygen partial pressure in RMS should safely rule out any oxygen deficiency. In turn, the dependence of $T_{\text{f}}$ on the growth rate, which is consistently found in \cite{YTK01} and Fig.\ \ref{fig3}(a), cannot depend on the existence of oxygen vacancies for RMS grown ZnFe$_{\text{2}}$O$_{\text{4}}$. To summarize this part, the two sample series shown in Fig.\ \ref{fig4} underline, that the relevant preparation parameter to control $T_{\text{f}}$ is the sputtering power and thus growth rate, while $T_{\text{growth}}$ and the Ar:O$_2$ ratio play a minor role in the resulting magnetic properties, in particular, $T_{\text{f}}$ is not directly controllable via oxygen vacancies. Therefore, in the following only samples grown at Ar:O$_2$ ratio of $10$\,:\,$0.5$ and $T_{\text{growth}} = $ \SI{450}{\degreeCelsius}, as those in Figs. \ref{fig1} to \ref{fig3}, will be discussed further.      

\begin{figure}[t]
	\centering
	\includegraphics[width=0.45\textwidth]{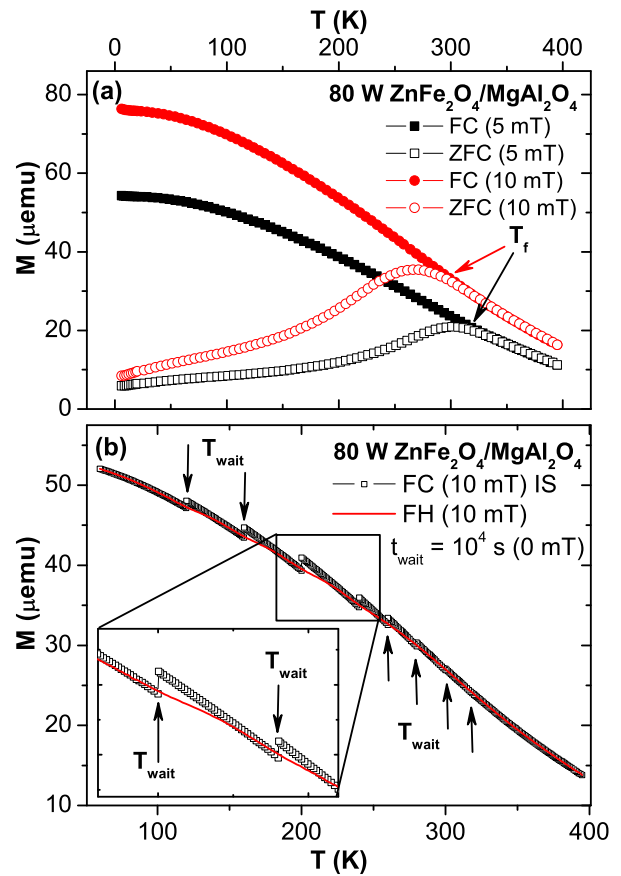}
	\vspace{-0.2cm}
	\caption{(a) $M(T)$ curves of the 80~W ZnFe$_{\text{2}}$O$_{\text{4}}$ grown on MgAl$_{\text{2}}$O$_{\text{4}}$ recorded under field cooled (FC, full symbols) and after zero-field cooled (ZFC, open symbols) conditions at 5~mT (black squares) and 10~mT (red circles). (b) $M(T)$ curves recorded while cooling at 10~mT (FC) with intermittent stops (IS, open squares) with various $T_{wait}$ with $t_{wait}$ of 10,000~s marked by arrows. The $M(T)$ curve while warming is subsequently recorded at 10~mT (red line).}
	\label{fig5}
\end{figure} 

In a next step, the actual type of magnetic order shall be determined because the reports in the literature range from ferro(i)magnetism, over superparamagnetism, to a cluster glass or spin-glass behavior. As pointed out above, $M_{\text{S}}$ for the present set of samples is found to be consistent with most of the reports found throughout the literature, while the bifurcation at $T_{\text{f}}$ is found at rather elevated temperatures. Figure \ref{fig5}(a) shows the $M(T)$ behavior recorded under FC conditions (full symbols) as well as after ZFC conditions (open symbols) for the ZnFe$_{\text{2}}$O$_{\text{4}}$ grown at \SI{80}{\watt} on MgAl$_{\text{2}}$O$_{\text{4}}$ for an external field of \SI{5}{mT} (black squares) and \SI{10}{mT} (red circles). As expected $T_{\text{f}}$ increases with decreasing external magnetic field which is consistent with both spin-freezing as well as superparamagnetism. Reducing the external field further to fields of 0.2 to 0.5~mT), where most of the spin glass experiments are typically carried out \cite{BeK09}, $T_{\text{f}}$ gets very close to the maximum attainable temperature of 400~K of the SQUID magnetometer (not shown). Therefore, all subsequent experiments are only carried out at fields of 5~mT or 10~mT to keep the maximum achievable temperature well above $T_{\text{f}}$. Note, that unfortunately the SQUID does not allow to go above the magnetic order temperature which is in all cases above 400~K. 

To get a first estimate on the existence of magnetic glassiness, the FC memory sequence used in \cite{CPL19} was carried out. Figure \ref{fig5}(b) shows the FC $M(T)$ curve recorded at 10~mT with intermittent stops (IS, open symbols) at various $T_{\text{wait}}$ which are marked with arrows. Here the field was reduced to 0~mT for a waiting time $t_{\text{wait}}$ of 10,000~s. Then the field was set to 10~mT again and the cooling down is resumed. Subsequently $M(T)$ is measured at 10~mT while heating at the same rate as during FC without any IS (FH, full line). In the FC curve clear steps can be seen for most $T_{\text{wait}}$ which are most pronounced just below the maximum of the ZFC curve in Fig.\ \ref{fig5}(a), i.\,e., also below $T_{\text{f}}$, while they are virtually absent above $T_{\text{f}}$. Note, that in Fig.\ \ref{fig5} we show the magnetic data in emu to demonstrate the absolute size of the steps in comparison to the detection limit of the SQUID of $2-4 \cdot 10^{-7}$~emu \cite{SSN11,BHH18}. These steps demonstrate magnetic relaxation during $t_{\text{wait}}$. More important, the subsequent FH curve shows clear inflection points around $T_{\text{wait}}$, and the inset enlarges the two most prominent ones. Therefore, there is a first experimental evidence for magnetic glassiness in epitaxial ZnFe$_{\text{2}}$O$_{\text{4}}$ analogous to ZnFe$_{\text{2}}$O$_{\text{4}}$ nanopowders in \cite{CPL19}; however, this glassiness extents to rather high temperatures which are only comparable to those reported in \cite{YTK01}. A caveat is still, that such a behavior can also be observed and modeled in superparamagnetic samples as discussed in detail in \cite{SJT05} where only subtleties in these type of FC memory sequences allow to distinguish a superparamagnet from a superspin-glass. Therefore ZFC memory experiments are needed in addition.

\begin{figure}[t]
	\centering
	\includegraphics[width=0.45\textwidth]{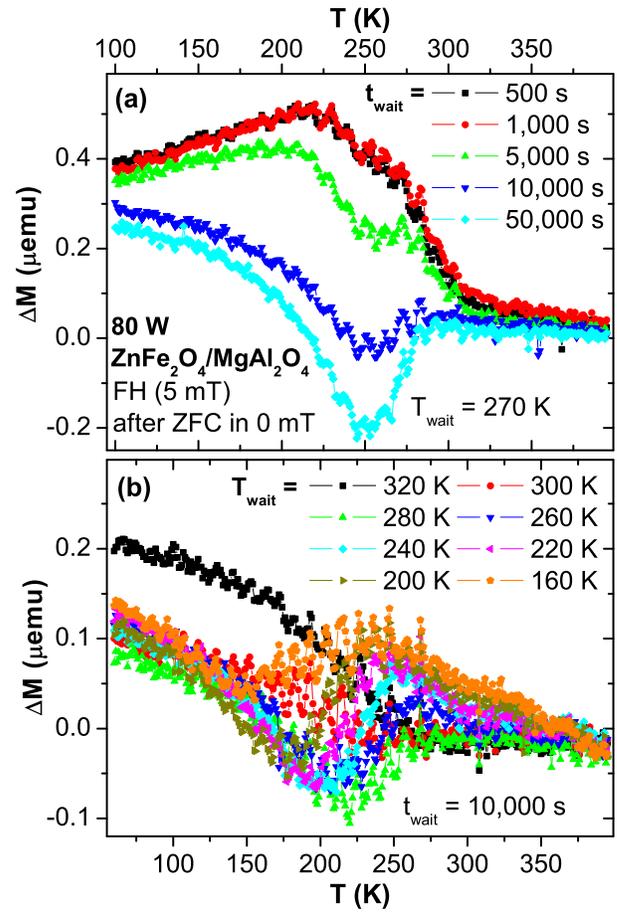}
	\vspace{-0.2cm}
	\caption{Characteristic hole-burning experiment for the 80~W ZnFe$_{\text{2}}$O$_{\text{4}}$ sample grown on MgAl$_{\text{2}}$O$_{\text{4}}$. (a) shows the dependence on the waiting time $t_{wait}$ for a fixed waiting temperature $T_{wait}$ of 270~K while (b) shows the dependence on $T_{wait}$ for a fixed $t_{wait}$ of 10,000~s.}
	\label{fig6}
\end{figure}

Figure \ref{fig6} provides additional experimental evidence for magnetic glassiness of the same sample by ZFC memory experiments adopted after \cite{BeK09}. Here the sample is cooled down under ZFC conditions once without any waiting time and once cooling is stopped at $T_{\text{wait}}=270$~K for varying waiting times $t_{\text{wait}}$ from 500~s to 50,000~s. Subsequently, an $M(T)$ curve is measured at 5~mT while warming (FH). In Fig.\ \ref{fig6}(a) the difference $\Delta M$ between the FH without and with $T_{\text{wait}}$ is plotted for all $t_{\text{wait}}$. Note, that $\Delta M$ is provided in emu and the visible scatter in the difference data is around $1 - 2 \cdot 10^{-8}$~emu, which demonstrates the high reproducibility of the data recorded with the SQUID magnetometer. It should be stressed that this is only possible if the magnet is reset before the measurement to eliminate any trapped flux. In all subsequent measurements one has to avoid magnetic fields larger than 10~mT so that the nominal and actual field are identical for all measurements within 0.1~mT between which $\Delta M$ is taken. For $t_{\text{wait}}$ of 500~s and 1,000~s an increase of $\Delta M$ below $T_{\text{f}}$ is visible with a maximum around the maximum of the ZFC $M(T)$ curve, i.\,e., $\Delta M$ follows the shape of the ZFC curve. However, the maximum of the ZFC curve for the given experimental conditions is around 300~K while the maximum of the $\Delta M$-curve is around 225~K, i.\,e., shifted to lower temperatures and does not go back to zero. This low temperature increase of $\Delta M$ is difficult to be explained in a straightforward manner, because the nominally ZFC conditions only correspond to less than 0.1~mT \cite{BHH18}. Therefore the difference between the ZFC with an without $t_{\text{wait}}$ of 500~s at 270~K, i.\,.e., below $T_{\text{f}}$ implies that the system is allowed to spend additional 500~s close to the freezing temperature in a tiny, but finite field. If one considers a superparamagnetic ensemble close to its blocking temperature this implies more time for thermally activated switching in a tiny field which imprints a tiny additional magnetization because the residual field induces a slight imbalance in the probability of switching parallel and antiparallel to it. A superparamagnetic ensemble would further imply relatively fast characteristic time-scales for the switching attempts. This would be in accordance that $\Delta M$ with $t_{\text{wait}}$ of 500~s and 1.000~s are virtually identical, because all the switching events are already done, while without $t_{\text{wait}}$ the system is ramped through the blocking temperature with a rate of 60~s/K, so much less switching events can take place around the blocking temperature, where the tiny residual field is sufficient to aid the thermally activated switching events. The remaining low temperature increase is thus the frozen-in result of more switching events close to $T_{\text{f}}$ resulting in a waiting-time imprinted additional magnetization. This is further corroborated by the fact, that the low-temperature increase is found to decrease with decreasing $T_{\text{wait}}$, i.\,e., a waiting further below $T_{\text{f}}$ and thus in a region with potentially slower dynamics, see Fig.\ \ref{fig6}(b). We thus infer that this low temperature increase of $\Delta M$ is most likely to be indicative of a superparamagnetic-like behavior with relatively fast dynamics rather than classical rejuvenation effects in superferromagnets as discussed in \cite{BeK09}. 

\begin{figure}[t]
	\centering
	\includegraphics[width=0.45\textwidth]{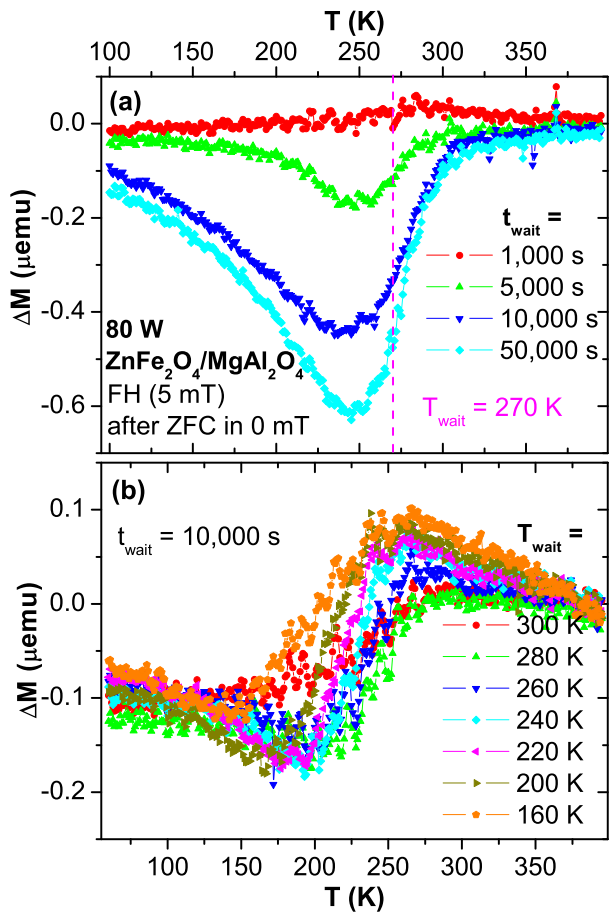}
	\vspace{-0.2cm}
	\caption{Identical set of data as in Fig.\ \ref{fig6} for the 80~W ZnFe$_{\text{2}}$O$_{\text{4}}$ sample grown on MgAl$_{\text{2}}$O$_{\text{4}}$ for (a) varying $t_{wait}$ for $T_{wait}=270$~K (magenta dahed line) and (b) varying $T_{wait}$ for $t_{wait}=$10,000~s; however, $\Delta M$ is taken differently (see text).}
	\label{fig6a}
\end{figure}

Beyond this low-temperature increase of $\Delta M$ seen for all $t_{\text{wait}}$ in Fig.\ \ref{fig6}(a), there is a minimum evolving with increasing $t_{\text{wait}}$ becoming clearly visible at 5,000~s and being most pronounced at 50,000~s. This is a typical characteristic of a (super)spin glass as discussed in \cite{BeK09,SJT05}; however, the minimum is shifted to lower temperatures compared to $T_{\text{wait}}$ by about 20~K. This shift is also seen, irrespective of the actual $T_{\text{wait}}$, see also Fig\ \ref{fig6a}(a) further below. Figure \ref{fig6}(b) shows $\Delta M$ curves for a fixed $t_{\text{wait}}$ of 10,000~s for various $T_{\text{wait}}$. For $T_{\text{wait}}$ above $T_{\text{f}}$ no minimum is visible and only the low temperature increase can be seen. In contrast, a clear minimum is observable which is strongest for $T_{\text{wait}}$ of 280~K, i.\,e., close to $T_{\text{f}}$. For lower $T_{\text{wait}}$ is becomes less pronounced and the minimum shifts to lower temperatures, which are however always below the respective $T_{\text{wait}}$, e.\,g., the minimum in $\Delta M$ for $T_{\text{wait}}$ of 160~K is at 150~K (orange pentagons). Therefore, the 80~W ZnFe$_{\text{2}}$O$_{\text{4}}$ sample grown on MgAl$_{\text{2}}$O$_{\text{4}}$ shows all the experimental characteristics of magnetic glassiness. On the one hand a FC memory effect characteristic of superparamagnets and (super)spin glasses \cite{SJT05} which have been reported for ZnFe$_{\text{2}}$O$_{\text{4}}$ before \cite{CPL19}, see Fig.\ \ref{fig5}(b). On the other hand, a $t_{\text{wait}}$-dependent minimum is observed, which is known as hole-burning experiment \cite{BeK09} and is absent in superparamagnets but is seen in (super)spin-glasses \cite{SJT05}. On the other hand, the low-temperature increase of $\Delta M$ for short $t_{\text{wait}}$ or $T_{\text{wait}}$ above $T_{\text{f}}$ resemble more of superparamagnetic-like behavior. However, we will show in the following that superparamagnetic-like behavior and magnetic glassiness coexist.

Figure \ref{fig6a} provides an alternative way of presenting the identical results as in Fig.\ \ref{fig6}. In this case $\Delta M$ is taken as the difference between all data with respect to (a) $t_{wait}=$500~s and (b) $T_{wait}=320$~K. In other words, here $\Delta M(T)$ should only contain the magnetic glassiness since the superparamagnetic behavior---which is reflected by the low temperature increase of $\Delta M$ seen for short $t_{wait}$ in Fig.\ \ref{fig6}(a), or $t_{wait}$ well above $T_{f}$ in Fig.\ \ref{fig6}(b)---is subtracted and thus only the slow, glassy dynamics can be seen. Figure \ref{fig6a}(a) reveals that $\Delta M(T)$ of $t_{wait}$ of 500~s and 1,000~s are virtually identical, since only a zero line is visible. In other words, the fast dynamics of the superparamagnet are over while the slow dynamics of the glassiness have not yet set in, both referring to the experimental accuracy. Having thus subtracted the fast dynamics the evolving dip in $\Delta M(T)$ with $t_{wait}$ of 5,000~s and higher nicely represents the remaining magnetic glassiness with its characteristic slow dynamics and ZFC memory effect. It is furthermore visible that the minimum in $\Delta M(T)$ does not align with $T_{wait}$ which is indicated by the dashed magenta line in Fig.\ \ref{fig6a}(a). $T_{wait}$ merely appears to align with the inflection point of the high-temperature side of $\Delta M(T)$ which is also seen in the $T_{wait}$-dependence of $\Delta M(T)$ in Fig.\ \ref{fig6a}(b). Also here, the low-temperature increase of $\Delta M$ seen in Fig.\ \ref{fig6}(b) is fully removed by taking $\Delta M$ always with respect to $T_{wait}=320$~K. Note, that in Fig.\ \ref{fig6a}(b) $T_{wait}=300$~K is not nicely visible but the dip in $\Delta M(T)$ is very weak and clearly less pronounced compared to the others and in fact may only reflect the limits of reproducibility of these types of SQUID experiments; one should keep in mind that two ZFC $M(T)$ curves like in Fig.\ \ref{fig5}(a) are subtracted from each other, i.\,e., the signal size and thus the relative accuracy of each data point varies (slightly) over the entire $T$-range which can easily affect difference signals of the order of $1 \cdot 10^{-7}$~emu. Nevertheless, Figs.\ \ref{fig6} and \ref{fig6a} nicely demonstrate that in ZnFe$_{\text{2}}$O$_{\text{4}}$ superparamagnetic and glassy behavior coexist and can be separated from each other. This is quite remarkable, since an epitaxial film of ZnFe$_{\text{2}}$O$_{\text{4}}$ is structurally quite distinct from a superparamagnetic ensemble like horse-spleen ferritin or a superspin-glass like a dense ensemble like Fe$_3$N nanoparticles which were both investigated in \cite{SJT05}. Yet, ZnFe$_{\text{2}}$O$_{\text{4}}$ epitaxial thin films exhibit both types of magnetic order at the same time. Therefore, the observed magnetic glassiness appears to be better described in terms of a cluster glass like in \cite{YTK01}, i.\,e., a superparamagnetic-like ensemble with (frustrated) intercluster interactions. However, these interactions have to be inhomogeneous and disordered throughout the sample and in contrast to the nanopowder in \cite{CPL19} they have no obvious structural origin. One has to therefore conclude that they stem from local variations of the cation distribution, i.\,e., from chemical or A/B disorder and thus they crucially depend on a finite amount of inversion. This in turn also explains why highly crystalline bulk ZnFe$_{\text{2}}$O$_{\text{4}}$ samples in \cite{HaC53,HaC56,KTK03} exhibit quite distinct magnetic properties.  

\begin{figure}[t]
	\centering
	\includegraphics[width=0.45\textwidth]{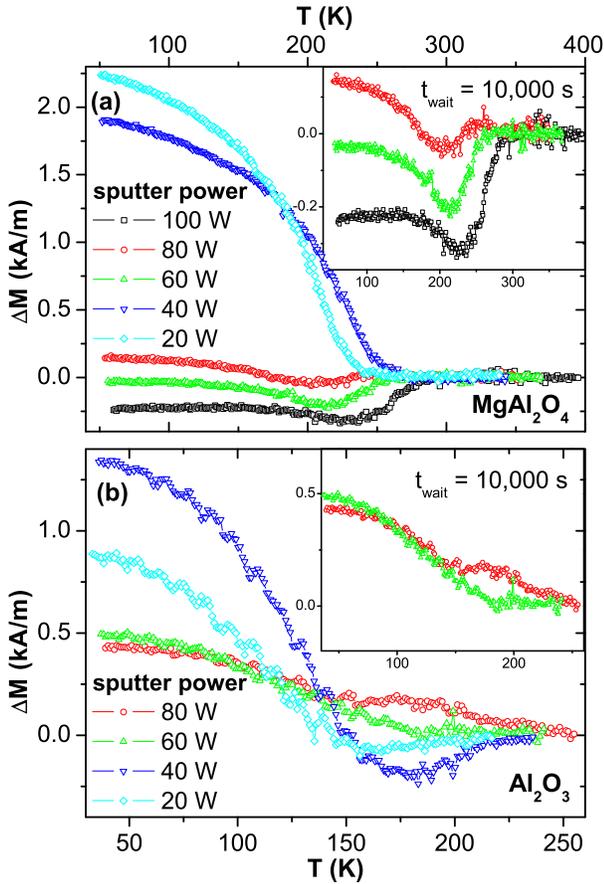}
	\vspace{-0.2cm}
	\caption{Comparison of the hole-burning experiments for ZnFe$_{\text{2}}$O$_{\text{4}}$ grown on MgAl$_{\text{2}}$O$_{\text{4}}$ (a) and Al$_{\text{2}}$O$_{\text{3}}$ (b) as a function of sputter power (details see text). The insets show the high sputter power samples only. }
	\label{fig7}
\end{figure} 

Since we have seen that $T_{\text{f}}$ is a function of the growth power during the sputtering process, the two power series of ZnFe$_{\text{2}}$O$_{\text{4}}$ samples grown on MgAl$_{\text{2}}$O$_{\text{4}}$ and Al$_{\text{2}}$O$_{\text{3}}$ shall be directly compared. For that we have chosen to perform the hole-burning ZFC waiting experiments of Fig.\ \ref{fig6}(b) on the identical relative temperature scale for each sample. In other words, the highest and lowest temperature of the $M(T)$ curves as well as  $T_{\text{wait}}$ have been chosen to be a the same relative temperature with respect to $T_{\text{f}}$ to assure that the samples spent comparable time-spans in regions with comparable magnetization dynamics. Note, that in addition the full experiment for all $T_{\text{wait}}$ of Fig.\ \ref{fig6}(b) on an absolute temperature scale have also been performed (not shown), but the direct comparison in essence reveals the identical result. Figure \ref{fig7} shows the $\Delta M$ curves for the power-series of ZnFe$_{\text{2}}$O$_{\text{4}}$ grown on MgAl$_{\text{2}}$O$_{\text{4}}$ (a) and Al$_{\text{2}}$O$_{\text{3}}$ (b) for $t_{\text{wait}}$ of 10,000~s; the insets enlarge the samples grown at high sputtering powers. Irrespective of the substrate the samples grown at sputtering powers of 20~W and 40~W do only show the low-temperature increase of $\Delta M$, i.\,e., mostly superparamagnetic-like behavior; for ZnFe$_{\text{2}}$O$_{\text{4}}$ on Al$_{\text{2}}$O$_{\text{3}}$ a faint and broad minimum is visible which however does not show a clear shift with $T_{\text{wait}}$ or a pronounced dependence with $t_{\text{wait}}$. Therefore, we consider this part as inconclusive, i.\,e., not as clear experimental evidence for glassiness. In contrast, the samples grown at 60~W and higher all show a hole-burning behavior in the ZFC memory experiments which is pronounced for ZnFe$_{\text{2}}$O$_{\text{4}}$ on MgAl$_{\text{2}}$O$_{\text{4}}$, see inset of Fig.\ \ref{fig7}(a), but rather weak for ZnFe$_{\text{2}}$O$_{\text{4}}$ on Al$_{\text{2}}$O$_{\text{3}}$, for which only a faint minimum can be seen, see inset of Fig.\ \ref{fig7}(b). Therefore, in ZnFe$_{\text{2}}$O$_{\text{4}}$ on Al$_{\text{2}}$O$_{\text{3}}$ only superparamagnetic-like can be inferred and signatures of magnetic glassiness are faint and limited to high sputtering powers. This goes hand-in-hand with a more pronounced maximum in the ZFC curves, see Fig.\ \ref{fig3}(b) and an increased magnetization, see Fig.\ \ref{fig2}(a). In contrast, ZnFe$_{\text{2}}$O$_{\text{4}}$ on MgAl$_{\text{2}}$O$_{\text{4}}$ exhibits a clear transition from superparamagnetic-like behavior at low sputtering powers with clear signs of magnetic glassiness existing at high sputtering powers, i.\,e., growth rates. To ultimately clarify what causes the discrepancy in the magnetic properties for ZnFe$_{\text{2}}$O$_{\text{4}}$ grown on MgAl$_{\text{2}}$O$_{\text{4}}$ and Al$_{\text{2}}$O$_{\text{3}}$ as well as at low and high sputtering powers, the 20~W and the 80~W samples were subjected to an element-selective magnetic characterization using XMCD.

\begin{figure}[t]
	\centering
	\includegraphics[width=0.45\textwidth]{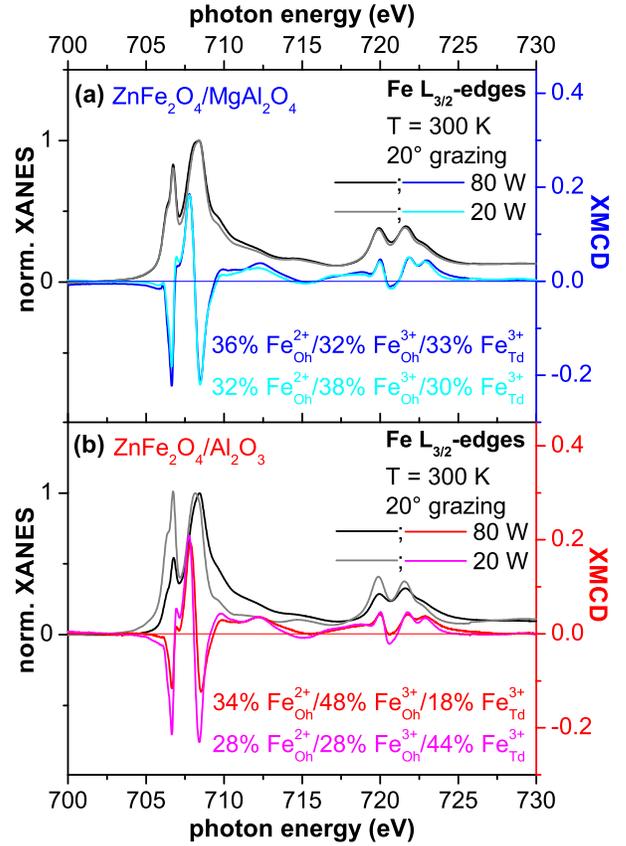}
	\vspace{-0.2cm}
	\caption{Normalized XANES and XMCD spectra recorded at the Fe $L_{3/2}$-edges under grazing incidence at 300~K for the 80~W and 20~W ZnFe$_{\text{2}}$O$_{\text{4}}$ samples grown on (a) MgAl$_{\text{2}}$O$_{\text{4}}$ and (b) Al$_{\text{2}}$O$_{\text{3}}$, respectively. The XMCD spectra have also been simulated to determine the relative amount of the individual Fe species (see text).}
	\label{fig8}
\end{figure}

Figure \ref{fig8} shows the measured XANES and XMCD spectra at the Fe $L_{3/2}$-edges for ZnFe$_{\text{2}}$O$_{\text{4}}$ grown on MgAl$_{\text{2}}$O$_{\text{4}}$ (a) and Al$_{\text{2}}$O$_{\text{3}}$ (b) for the samples grown at 20~W and 80~W, respectively. The XMCD at the Fe L$_3$-edge has been also simulated by respective multiplet ligand field theory using the CTM4XAS code using the identical parameters as in \cite{LBP20}. In brief, the negative peaks in the XMCD spectrum are stemming from the octahedral contributions Fe$_{Oh}$, where Fe$^{2+}_{Oh}$ is mostly seen at lower (706.6~eV) and Fe$^{3+}_{Oh}$ at higher (708.5~eV) photon energies; the positive peak at 707.8~eV can be assigned to Fe$^{3+}_{Td}$. The experimental XMCD can be reproduced by adjusting the relative concentrations of Fe$^{3+}_{Oh}$, Fe$^{2+}_{Oh}$, and Fe$^{3+}_{Td}$ to match the experimental XMCD; the results of this are given in Fig.\ \ref{fig8}. It can be seen in Fig.\ \ref{fig8}(a), that there are no pronounced differences between the experimental XMCD spectra of the Fe $L_{3/2}$-edge XMCD for ZnFe$_{\text{2}}$O$_{\text{4}}$/MgAl$_{\text{2}}$O$_{\text{4}}$ grown at either 20~W or 80~W as well as for the respective results of the simulation. About one third of the Fe is located on tetrahedral sites, i.\,e., the degree of inversion $\delta$ is around 0.3 for both sputtering powers. Also a significant amount of Fe$^{2+}_{Oh}$ is found, which would suggest a strong contribution from a $J_{\text{BB}}^{\text{DE}}$ double exchange interaction which appears to be slightly larger for the 80~W sample which exhibits the magnetic glassiness in comparison to the 20~W sample, which only shows the superparamagnetic-like low temperature increase of $\Delta M$.  The ZnFe$_{\text{2}}$O$_{\text{4}}$/Al$_{\text{2}}$O$_{\text{3}}$ samples in Fig.\ \ref{fig8}(b) exhibit a different behavior. Here the 80~W sample has a strongly reduced contribution of Fe$^{3+}_{Td}$ compared to the $Fe_{Oh}$ compared to the 20~W sample which has the highest relative content of Fe$^{3+}_{Td}$. On the other hand, the actual positive peak in the XMCD is of identical size in both samples. It therefore appears, the XMCD intensity for the $Fe_{Oh}$ is reduced while the amount of Fe$^{3+}_{Td}$ remains constant. This may appear as contradiction at first sight, since the relative contents may suggest different degrees of inversion for the two samples. However, one should keep in mind that the magnetic superexchange interaction on the octahedral sites $J_{\text{BB}}$ is weakly antiferromagnetic while double exchange leads to spin canting \cite{VAN09,ZGS20}. Since the magnetic order is observed up to above room temperature for all samples in this work, the $J_{\text{AB}}$ superexchange mechanism has to play a significant role, which is consistent with a finite degree of inversion of the order of 0.3. In that light, the presence of a finite amount of inversion giving rise to Fe$^{3+}_{Td}$ is a prerequisite for magnetic order at elevated temperatures but does not play a decisive role for the presence of magnetic glassiness, since Fe$^{3+}_{Td}$ is found in all four samples while glassiness is only found in the 80~W samples, in particular in those grown on MgAl$_{\text{2}}$O$_{\text{4}}$. In addition, the presence of Fe$^{2+}_{Oh}$ in all samples further suggests the presence of an additional $J_{\text{BB}}^{\text{DE}}$ double exchange mechanism associated with spin canting. Here the relative amount of Fe$^{2+}_{Oh}$ increases only slightly from the 20~W sample on Al$_{\text{2}}$O$_{\text{3}}$ over 20~W on MgAl$_{\text{2}}$O$_{\text{4}}$, 80~W on Al$_{\text{2}}$O$_{\text{3}}$ to 80~W on MgAl$_{\text{2}}$O$_{\text{4}}$, i.\,e., it follows the trend of increasing glassiness of the samples. However, the changes are rather small and the significance of determining such small changes with multiplet ligand field simulations is limited. Obviously, there is no straightforward mechanism for the occurrence of magnetic glassiness which can be derived form the XMCD spectra at the Fe $L_{3/2}$-edges. A finite degree of inversion has to play a role but mostly for the high order temperatures observed. The existence of glassiness appears to be linked to a delicate balance of the various competing exchange interactions as well as the local cationic configuration which has to be inhomogeneous throughout the sample as discussed above. Obviously the growth rate influences mostly the latter where the highest growth rates favor glassiness, most likely via increased local cationic disorder, in particlular in ZnFe$_{\text{2}}$O$_{\text{4}}$ on MgAl$_{\text{2}}$O$_{\text{4}}$ substrates.   

\begin{figure}[t]
	\centering
	\includegraphics[width=0.45\textwidth]{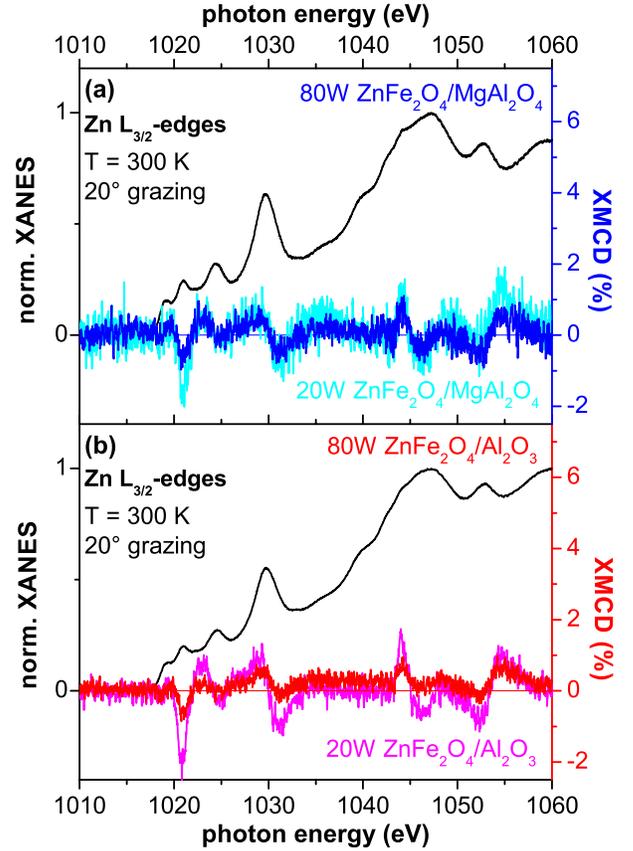}
	\vspace{-0.2cm}
	\caption{Normalized XANES and XMCD spectra recorded at the Zn $L_{3/2}$-edges under grazing incidence at 300~K for the 80~W and 20~W ZnFe$_{\text{2}}$O$_{\text{4}}$ samples grown on (a) MgAl$_{\text{2}}$O$_{\text{4}}$ and (b) Al$_{\text{2}}$O$_{\text{3}}$, respectively.}
	\label{fig9}
\end{figure} 

Figure \ref{fig9} shows the measured XANES and XMCD spectra at the Zn $L_{3/2}$-edges of the identical set of samples as in Fig.\ \ref{fig8}. The XANES for ZnFe$_{\text{2}}$O$_{\text{4}}$ on MgAl$_{\text{2}}$O$_{\text{4}}$ in Fig.\ \ref{fig9}(a) is rather similar to the one of ZnFe$_{\text{2}}$O$_{\text{4}}$ on Al$_{\text{2}}$O$_{\text{3}}$ in (b). All four samples exhibit a finite XMCD with comparable spectral shape; all XMCD spectra were derived by reversing both, helicity of the light as well as the magnetic field and it was verified that the XMCD spectrum nicely reverses with reversing external field (not shown). The size of the Zn $L_{3/2}$-edge XMCD follows the amount of Fe$^{3+}_{Td}$ as seen in Fig.\ \ref{fig8}, i.\,e., the Zn XMCD is largest for the 20~W sample on Al$_{\text{2}}$O$_{\text{3}}$, which has the highest relative Fe$^{3+}_{Td}$ content and it is lowest for the 20~W sample on Al$_{\text{2}}$O$_{\text{3}}$ which has the lowest relative Fe$^{3+}_{Td}$ content. It is thus reasonable to assume that the magnetic polarization of Zn in ZnFe$_{\text{2}}$O$_{\text{4}}$ is mostly associated with Zn$^{2+}_{Oh}$. In turn, this implies that a weakly polarized cation substitutes for a strongly polarized one thus reducing the effective exchange. This is consistent with the experimental observation that the 80~W ZnFe$_{\text{2}}$O$_{\text{4}}$/MgAl$_{\text{2}}$O$_{\text{4}}$ has the highest $T_{\text{f}}$ and the lowest magnetic polarization of Zn while the highest Zn polarization in the 20~W ZnFe$_{\text{2}}$O$_{\text{4}}$/Al$_{\text{2}}$O$_{\text{3}}$ sample is associated with the lowest $T_{\text{f}}$. To verify this hypothesis, more sophisticated theoretical calculations beyond the multiplet ligand field codes is required where the individual spectroscopic signatures in the Zn $L_{3/2}$-edge XANES and XMCD can be associated with the actual Zn species which however goes beyond the scope of this paper. Nevertheless, it is already evident that a too high degree of inversion as seen by strong magnetic polarization of the Zn together with a high relative content of Fe$^{3+}_{Td}$ is unfavorable for both, high $T_{\text{f}}$ as well as magnetic glassiness and high growth rates appear to be an experimental means to control/limit excessive inversion but at the same time assure sufficient local cationic disorder to induce magnetic glassiness. 

\section{Discussion and conclusion}

ZnFe$_{\text{2}}$O$_{\text{4}}$ epitaxial thin films have been grown on MgAl$_{\text{2}}$O$_{\text{4}}$ and Al$_{\text{2}}$O$_{\text{3}}$ substrates with varying preparation conditions. All samples were investigated with respect to their basic structural and magnetic properties and long range magnetic order was found above room temperature for all samples. The stoichiometric composition of the samples was verified using RBS. A clear bifurcation between $M(T)$ curves under FC and ZFC conditions is found at $T_{\text{f}}$, which is systematically higher for ZnFe$_{\text{2}}$O$_{\text{4}}$ on MgAl$_{\text{2}}$O$_{\text{4}}$ by about 100~K. $T_{\text{f}}$ is found to systematically increase with increasing the sputtering power and thus growth rate in agreement with \cite{YTK01}. The Ar:O$_2$ ratio was not found to influence neither $T_{\text{f}}$ nor $M_s$ in a systematic manner; increasing $T_{\text{growth}}$ increases only $M_s$ while $T_{\text{f}}$ exhibits no systematic changes. 

An in-depth study of the magnetic properties using FC as well as ZFC memory experiments reveals magnetic glassiness for samples grown at high sputter powers. The glassiness is more pronounced for ZnFe$_{\text{2}}$O$_{\text{4}}$/MgAl$_{\text{2}}$O$_{\text{4}}$ compared to ZnFe$_{\text{2}}$O$_{\text{4}}$/Al$_{\text{2}}$O$_{\text{3}}$, where the signatures of magnetic glassiness beyond those in FC memory experiments are generally weak. At lower growth rates the signatures of glassiness are absent and a low-temperature increase of $\Delta M$ is observed which points towards superparamagnetic-like behavior and the signatures in FC memory experiments are weak and the ZFC memory experiments shows no hole burning effect in accordance with the expectations for superparamagnetic samples \cite{SJT05}. In contrast, at high growth rates, in particular for ZnFe$_{\text{2}}$O$_{\text{4}}$/MgAl$_{\text{2}}$O$_{\text{4}}$ ZFC memory experiments show an additional hole burning effect which is characteristic for spin glasses \cite{BeK09} and superspin glasses \cite{SJT05}. Since the glassiness coexists with superparamagnetic-like signatures, in particular, the low temperature increase of $\Delta M$, the structural properties of the ZnFe$_{\text{2}}$O$_{\text{4}}$ epitaxial films are quite different from the nanoparticle ensembles in \cite{SJT05} the observed magnetic properties are described best as cluster glass in analogy to comparable observations for epitaxial ZnFe$_{\text{2}}$O$_{\text{4}}$ in \cite{YTK01}. 

An in-depth characterization based on XANES and XMCD reveals that a finite magnetic polarization at the Zn $L_{3/2}$ edges exists in all ZnFe$_{\text{2}}$O$_{\text{4}}$ samples which adds more complexity to the magnetic interactions beyond the usually discussed Fe-based exchange. At the Fe $L_{3/2}$ the XMCD is used to extract the relative concentrations of Fe$^{3+}_{Td}$, Fe$^{3+}_{Oh}$, and Fe$^{2+}_{Oh}$ by means of multiplet ligand field simulations as done before \cite{KMK14,LBP20}. The abundance of Fe$^{3+}_{Td}$ correlates well with the size of the magnetic polarization of Zn and thus both can serve as a measure for the degree of inversion. For highest inversion $T_{\text{f}}$ is found to be lowest and signatures of magnetic glassiness are absent. In contrast, the sample with the strongest signatures of magnetic glassiness, the 80~W ZnFe$_{\text{2}}$O$_{\text{4}}$/MgAl$_{\text{2}}$O$_{\text{4}}$, exhibits no significant changes in the Fe $L_{3/2}$-edge XMCD compared to the superparamagnetic-like 20~W sample. The most prominent tendency appears to be the relative amount of Fe$^{2+}_{Oh}$ which can be associated with a double-exchange mechanism which was held responsible for spin canting \cite{VAN09,ZGS20}. Since the differences in XMCD between the respective samples are small, this correlation between glassiness and Fe$^{2+}_{Oh}$ cannot be taken as significant but merely a starting point for more elaborate theoretical work to understand the details of the obtained XMCD spectra beyond the multiplet ligand field simulations. Most likely the cluster glass behavior in epitaxial ZnFe$_{\text{2}}$O$_{\text{4}}$ cannot be assigned to the actual structure of the materials like in common superparamagnets or dense nanoparticle ensembles \cite{SJT05} but due to local variations of the stoichiometry, leading to an inhomogeneous local cation distribution throughout the sample. As a consequence, the local magnetic moments in ZnFe$_{\text{2}}$O$_{\text{4}}$ are disordered due to partial inversion and partially canted due to the presence of Fe$^{2+}_{Oh}$, which leads to characteristic signatures of a cluster glass at rather high temperatures which is mostly controllable by the growth rate as reported for ZnFe$_{\text{2}}$O$_{\text{4}}$ before \cite{YTK01}. 

\section*{Acknowledgments}

J. L. gratefully acknowledges funding by FWF project ORD-49 at the initial stage of this work. A.Z. acknowledges the financial support by the Swiss National Science Foundation (SNSF) under Project No. 200021-169467. The X-ray absorption measurements were performed on the EPFL/PSI X-Treme beamline at the Swiss Light Source, Paul Scherrer Institut, Villigen, Switzerland. In addition, support by VR-RFI (Contracts No. 2017-00646 9 and No. 2019-00191) and the Swedish Foundation for Strategic Research (SSF, Contract No. RIF14-0053) supporting accelerator operation at Uppsala University is gratefully acknowledged. The research leading to this result has been supported by the RADIATE project under the Grant Agreement No. 824096 from the EU Research and Innovation programme HORIZON 2020.

\end{document}